\documentclass[12pt]{elsart3p}

\usepackage{color,times}
\usepackage{amssymb,amsmath,graphicx,color,url}
\usepackage{lineno}
\usepackage{natbib}
\usepackage{textcomp}

\journal{Chemical Engineering Science} 

\begin{document}
\begin{frontmatter}
\title{Multifractal detrended fluctuation analysis of combustion flames in four-burner impinging entrained-flow gasifier}
\author[KL,ICCT]{Miao-Ren Niu},
\author[KL,ECUST]{Wei-Xing Zhou\corauthref{cor}},
\corauth[cor]{Corresponding author. Address: 130 Meilong Road, P.O.
Box 114, East China University of Science and Technology, Shanghai
200237, China, Phone: +86 21 64253634, Fax: +86 21 64253152.}
\ead{wxzhou@ecust.edu.cn} %
\author[KL,ICCT]{Zhuo-Yong Yan},
\author[KL,ICCT]{Qing-Hua Guo},
\author[KL,ICCT]{Qin-Feng Liang},
\author[KL,ICCT]{Fu-Chen Wang},
\author[KL,ICCT]{Zun-Hong Yu}
\address[KL]{Key Laboratory of Coal Gasification of Ministry of Education, East China University of Science and Technology, Shanghai 200237, China}
\address[ICCT]{Institute of Clean Coal Technology, East China University of Science and Technology, Shanghai 200237, China}
\address[ECUST]{School of Business, School of Science, Research Center for Econophysics, and Research
Center of Systems Engineering, East China University of Science and
Technology, Shanghai 200237, China}

\begin{abstract}
On a laboratory-scale testing platform of impinging entrained-flow
gasifier with four opposed burners, the flame images for diesel
combustion and gasification process were measured with a single
charge coupled device (CCD) camera. The two-dimensional multifractal
detrended fluctuation analysis was employed to investigate the
multifractal nature of the flame images. Sound power-law scaling in
the annealed average of detrended fluctuations was unveiled when the
order $q>0$ and the multifractal feature of flame images were
confirmed. Further analyses identified two multifractal parameters,
the minimum and maximum singularity $\alpha_{\min}$ and
$\alpha_{\max}$, serving as characteristic parameters of the
multifractal flames. These two characteristic multifractal
parameters vary with respect to different experimental conditions.
\end{abstract}


\end{frontmatter}

\typeout{SET RUN AUTHOR to \@runauthor}

\section{Introduction}

Coal gasification is an important chemical process for coal and the
key to realize coal clean utilization. Entrained flow gasification
is leading among coal gasification technologies and represents one
of the cleanest ways of coal utilization. The entrained-flow
gasification technology has been extensively applied to the
production of ammonia, methanol, acetic acid, other chemicals and
power generation through Integrated Gasification Combined Cycle
($\rm IGCC$). The gasification process of an entrained-flow gasifier
is very complicated, because it relates to the fluid flow under the
condition of high temperature, high pressure and heterogeneous
state. Impinging stream flow configurations are characterized by
streams of fluid jets impinging against each other in a confined
vessel, which have proved useful in conducting a wide array of
chemical engineering unit operations and enhancing heat and mass
transfer between phases due to the high transfer coefficients
\citep{Tamir-1994}. The opposing jet technique has been applied in
many fields and extensively studied both practically
\citep{Tamir-Elperin-Luzzatto-1984-CES,Nosseir-Behart-1986-AIAAJ,Berman-Tamir-1996-CJCE,Berman-Tanklevsky-Oren-Tamir-2000a-CES,Berman-Tanklevsky-Oren-Tamir-2000b-CES,Dehkordi-2002-CEP}
and theoretically
\citep{Champion-Libby-1993-PFA,Kostiuk-Libby-1993-PFA}.

Fractals and multifractals are ubiquitous in natural and social
sciences \citep{Mandelbrot-1983}. The fractal behaviour of turbulent
premixed flame fronts has been manifested in many experiments
\citep{Gouldin-1987-CF,Gouldin-Hilton-Lamb-1989-PCI,Murayama-Takeno-1989-PCI,Mantzaras-Felton-Bracco-1989-CF,Goix-Shepherd-Trinite-1989-CST,Shepherd-Cheng-Goix-1991-PCI,North-Santavicca-1990-CST,Wu-Kwon-Driscoll-Faeth-1991-CST,Goix-Shepherd-1993-CST,Yoshida-Ando-Yanagisawa-Tsuji-1994-CST,Yoshida-Kasahara-Tsuji-Yanagisawa-1994-CST,Smallwood-Gulder-Snelling-Deschamps-Gokalp-1995-CF,Erard-Boukhalfa-Puechberty-Trinite-1996-CST,Das-Evans-1997-EF}.
These studies provided a quantitative description of flames, which
enables us to better understand the dynamics of different types of
combustion. On the other hand, to the best of our knowledge, the
multifractal nature of impinging flames has not been studied. We
shall adopt the two-dimensional multifractal detrended fluctuation
analysis (MFDFA) recently developed by \citet{Gu-Zhou-2006-PRE} for
this purpose. The 2D MFDFA is a generalization of the 1D DFA and
MFDFA invented by
\citet{Peng-Buldyrev-Havlin-Simons-Stanley-Goldberger-1994-PRE} and
\citet{Kantelhardt-Zschiegner-Bunde-Havlin-Bunde-Stanley-2002-PA},
which is widely used to investigate fractal behavior of time series.
The family of DFA approaches have the advantages of easy
implementation, high precision, and low computational time
\citep{Peng-Buldyrev-Havlin-Simons-Stanley-Goldberger-1994-PRE,Kantelhardt-Zschiegner-Bunde-Havlin-Bunde-Stanley-2002-PA,Gu-Zhou-2006-PRE}.

The paper is organized as follows. Section \ref{s1:Method} reviews
the algorithm of the two-dimensional MFDFA in detail. Under each
experimental condition, we have recorded many flame images. Rather
than analyze the images one by one under the same experimental
condition, we propose to perform annealed averaging over dozens of
images. The method of annealed averaging is also described. Section
\ref{s1:Experiments} outlines the schematic diagram of the
experiment setup and the details of the experiments. Section
\ref{s1:results} performs multifractal DFA analysis of the flame
images, investigates the dependence of multifractal characteristics
with respect to experimental conditions, and discusses the physical
interpretation of multifractality. Section \ref{s1:Concl}
summarizes.

\section{Methodology}
\label{s1:Method}

\subsection{Two-dimensional MFDFA}

The idea of DFA was invented originally by
\citet{Peng-Buldyrev-Havlin-Simons-Stanley-Goldberger-1994-PRE} to
investigate the long-range dependence in coding and noncoding DNA
nucleotide sequences and then generalized by
\citet{Kantelhardt-Zschiegner-Bunde-Havlin-Bunde-Stanley-2002-PA} to
study the multifractal nature hidden in time series, termed as
multifractal DFA (MFDFA). Recently, \citet{Gu-Zhou-2006-PRE}
generalized the one-dimensional MFDFA to higher-dimensional version,
which is capable of analyzing multifractal properties of
higher-dimensional objects. According to \citet{Gu-Zhou-2006-PRE},
the two-dimensional MFDFA consists of the following steps.

Step 1: Consider a flame image $t$, which is denoted by a
two-dimensional array $X(t; i,j)$, where $i=1,2,\cdots,M$, $j = 1,
2,\cdots, N$, and $t=1,2, \cdots, T$. The surface is partitioned
into $M_sN_s$ disjoint square segments of the same size $s\times s$,
where $M_s = [M/s]$ and $N_s = [N/s]$. Each segment can be denoted
by $X_{v,w}(t)$ such that $X_{v,w}(t;i,j)=X(t;l_1+i,l_2+j)$ for
$1\leqslant{i,j}\leqslant{s}$, where $l_1=(v-1)s$ and $l_2=(w-1)s$.

Step 2: For each segment $X_{v,w}(t)$, the cumulative sum
$u_{v,w}(t;i,j)$ is calculated as follows:
\begin{equation}
u_{v,w}(t;i,j) =
\sum_{k_1=1}^{i}\sum_{k_2=1}^{j}{X_{v,w}(t;k_1,k_2)}~,
 \label{Eq:eq1}
\end{equation}
where $1\leqslant{i,j}\leqslant{s}$. Note that $u_{v,w}$ itself is a
surface.

Step 3: The trend of the constructed surface $u_{v,w}(t)$ can be
determined by fitting it with a prechosen bivariate polynomial
function $\widetilde{u}$. In this work, the following polynomial is
adopted,
\begin{equation}
\widetilde{u}_{v,w}(t; i,j)=ai^2+bj^2+cij+di+ej+f~,\label{Eq:eq2}
\end{equation}
where $1\leqslant{i,j}\leqslant{s}$, and $a$, $b$, $c$, $d$, $e$,
and $f$ are free parameters to be determined. These parameters can
be estimated easily through simple matrix operations, derived from
the least squares method. We can then obtain the residual matrix
\begin{equation}
\epsilon_{v,w}(t;i,j)=u_{v,w}(t;i,j)-\widetilde{u}_{v,w}(t;i,j)~.
\label{Eq:eq3}
\end{equation}
The detrended fluctuation function $F(t; v,w,s)$ of the segment
$X_{v,w}(t)$ is defined via the sample variance of the residual
matrix $\epsilon_{v,w}(t;i,j)$ as follows
\begin{equation}
F^2(t; v,w,s) = \frac{1}{s^2}\sum_{i = 1}^{s}\sum_{j =
1}^{s}\epsilon_{v,w}(t; i,j)^2~. \label{Eq:eq4}
\end{equation}

Step 4: The overall detrended fluctuation is calculated by averaging
over all the segments, that is,
\begin{equation}
F_q(t;s) = \left\{\frac{1}{M_sN_s}\sum_{v = 1}^{M_s}\sum_{w =
1}^{N_s}{[F(t;v,w,s)]^q}\right\}^{1/q}, \label{Eq:eq5}
\end{equation}
where $q$ can take any real value except for $q = 0$. When $q = 0$,
we have
\begin{equation}
F_0(t;s) = \exp\left\{\frac{1}{M_sN_s}\sum_{v = 1}^{M_s}\sum_{w =
1}^{N_s}{\ln[F(t;v,w,s)]}\right\}~, \label{Eq:eq6}
\end{equation}
according to L'H\^{o}pital's Rule.

Step 5: Varying the value of $s$ in the range from $s_{\min} \approx
6$ to $s_{\max} \approx \min(M,N)/4$, we can determine the scaling
relation between the detrended fluctuation function $F_q(t;s)$ and
the size scale $s$, which reads
\begin{equation}
F_q(t;s) \sim s^{h(q)}~. \label{Eq:eq7}
\end{equation}
Usually, the scaling laws hold in a properly determined scaling
range.

Since $M$ and $N$ are often not a multiple of the segment size $s$,
two orthogonal strips at the end of the profile may remain. In order
to take these ending parts of the surface into consideration, the
same partitioning procedure can be repeated starting from the other
three corners. In this way, we have $4M_sN_s$ segments and the
overall detrended  fluctuation is calculated by averaging over them.

The final outcome of the MFDFA analysis is a family of scaling
exponents $h(q)$ which is a decreasing function of $q$ for
multifractal surface and remains constant for monofractals. In the
standard multifractal formalism of
\citet{Halsey-Jensen-Kadanoff-Procaccia-Shraiman-1986-PRA} based on
partition functions, the multifractal nature is characterized by the
mass exponent ${\tau(q)}$, which is a nonlinear function of $q$ .
\citet{Kantelhardt-Zschiegner-Bunde-Havlin-Bunde-Stanley-2002-PA}
showed that, for each $q$, we can obtain the corresponding
traditional ${\tau(q)}$ function through
\begin{equation}
\tau(q) = qh(q) - D_f~, \label{Eq:eq8}
\end{equation}
where $D_f$ is the fractal dimension of the geometric support of the
multifractal measure. In this work, we have $D_f=2$. Resorting to
the Legendre transform
\citep{Halsey-Jensen-Kadanoff-Procaccia-Shraiman-1986-PRA}, We can
also determine the singularity strength function ${\alpha(q)}$ and
the multifractal spectrum $f({\alpha})$ as follows
\begin{equation}
\alpha(q) = h(q) + qh'(q)~, \label{Eq:eq9}
\end{equation}
\begin{equation}
f({\alpha}) = q\alpha(q) - \tau(q)~. \label{Eq:eq10}
\end{equation}

\subsection{Annealed averaging}

To achieve better accuracy and higher statistical significance, we
perform annealed averaging over 50 images for each experimental
condition, which are selected randomly from a huge data base. The
annealed averaging gives the means of $F_{q}(s)$ over the selected
images,
\begin{equation}
F_q(s) =\left\{\frac{1}{T}\sum_{t=1}^T[F_q(t;s)]^q\right\}^{1/q}~.
\end{equation}
When $q = 0$, $F_0(s)$ can be calculated according to the following
expression
\begin{equation}
\ln[F_0(s)]= \frac{1}{T}\sum_{t=1}^T\ln[F_0(t;s)]~. \label{Eq:eq13}
\end{equation}
The characteristic functions of the ensemble of multifractal images
can be determined according to Eqs.~(\ref{Eq:eq7}-\ref{Eq:eq10}).

The shape and width of $f({\alpha})$ curve contain significant
information about the singular flame images. In general, the
spectrum has a concave downward curvature. We obtain two
characteristic quantities, the width of multifractal spectrum
$\Delta\alpha=\alpha_{\max} - \alpha_{\min}$ and the difference of
fractal dimensions $\Delta f = f(\alpha_{\max}) - f(\alpha_{\min})$
of the minimum probability subset with ${\alpha = \alpha_{\max}}$
and the maximum subset with ${\alpha = \alpha_{\min}}$. These two
quantities are used in this work to characterize different turbulent
flames under different conditions.

\section{Experimental equipment and procedure}
\label{s1:Experiments}

\begin{figure}[htp]
\centering
\includegraphics[width=6.5cm]{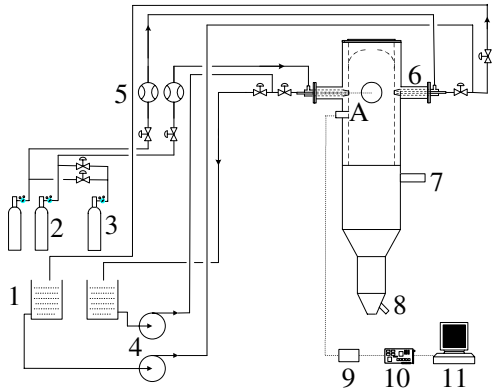}
\caption{Schematic diagram of experimental apparatus: 1$-$Liquid
tank, 2$-$$\rm O_2$ steel cylinder, 3$-$$\rm N_2$ steel cylinder,
4$-$Pump, 5$-$Gas mass flow meter, 6$-$Burner, 7$-$Vortex flowmeter,
8$-$Slag discharge, 9$-$Flame monitoring system.} \label{Fig1}
\end{figure}

The schematic drawing of the experimental apparatus is shown in
Fig.~\ref{Fig1}. The maximum values of the operation pressure and
temperature were 1 MPa and $1500^{\circ}\mathrm{C}$, respectively.
The gasifier was cylindrical, vertically oriented. The combustion
chamber was composed of a 15 mm thick cast refractory shell and its
inner diameter and length were 300 and 2200 mm, respectively. The
cast refractory shell, wrapped with a 235 mm thick and low thermal
conductivity fiber blanket to reduce the heat transfer, was
protected by a stainless-steel column shell of 0.8 m in diameter and
2.5 m in height. Ports were located on the wall of the gasifier for
viewing, temperature measurement, and insertion of a water-cooled
probe. Opposed turbulent flow fields were obtained by four opposed
round burners composed of inner and outer channels. Oxygen was fed
into the outer channels of burners by steel cylinder, with a
pressure-reducing valve to avoid pressure oscillations and achieve
steady flow. The gas flow rates were measured by mass flow meters
(D07-9C/ZM, Beijing Sevenstar Huachuang Electronic Co., Ltd). Diesel
oil was fed into the inner channels of burners by a gear pump
(A-73004-00 ${\rm \sharp}$, America Cole-Parmer Company), whose flux
was determined gravimetrically with an elapsed timer and an
electronic weight scale.

\begin{figure}[htp]
\centering
\includegraphics[width=6.5cm]{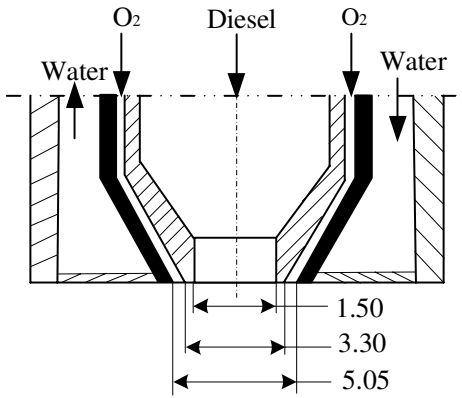}
\caption{General view of the burner (all dimensions in mm)}
\label{Fig2}
\end{figure}

In the gasification process, four burners were used to produce
opposite jets of fuel that impinge at the center of the combustion
chamber. The size of the burners is shown in Fig.~\ref{Fig2}. High
relative velocities between the particulate matter and the gaseous
phase in the central area provided good conditions for active
diffusion and convection at the particle surface, and the high
temperature together resulted in fast burning and gasification
reaction under highly reducing conditions to produce raw syngas.
High-temperature gaskets interfaced the furnace segments and
eliminated all leakage. From the reaction chamber, the raw syngas
flowed into the quench chamber, where the raw syngas was cooled and
partially scrubbed by water and then discharged. The impinging
flames are recorded with a flame monitoring system, fixed on the top
of the gasifier. After each experiment, nitrogen was fed into inner
channels by a steel cylinder to clean the burners.

\begin{figure}[htp]
\centering
\includegraphics[width=6.5cm]{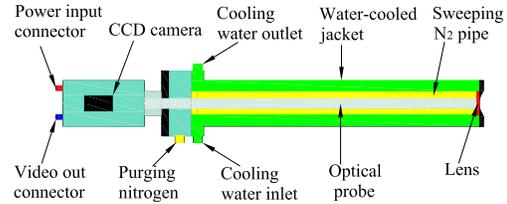}
\caption{(Color online) Schematic structure of the flame image
detector} \label{Fig3}
\end{figure}

The schematic structure of flame image detector is shown in
Fig.~\ref{Fig3}. The flame image detector is fixed on the top
refractory wall of the gasifier, and the whole flames are visible
during the experimental process. The flame monitoring system
consists of a lens, an optical probe, a CCD camera (Panasonic
WV-CP470), and a microcomputer. The CCD camera and its accessories
are cooled by water to avoid overheating. The objective lens is
fixed at the front end of the optical probe and its surface is kept
clean against dusts by a jet of nitrogen. The light conveyed by the
optical probe enters the CCD camera. The camera has a 1/3-inch high
resolution progressive scan interline-transfer CCD sensor with an
array of 720 ${\times}$ 576 pixels and each pixel contains 3 bytes
(one for each of the fundamental colors: red, green and blue). The
video signal is transferred from the CCD camera to the microcomputer
and stored.

\begin{figure}[htp]
\centering
\includegraphics[width=5.5cm]{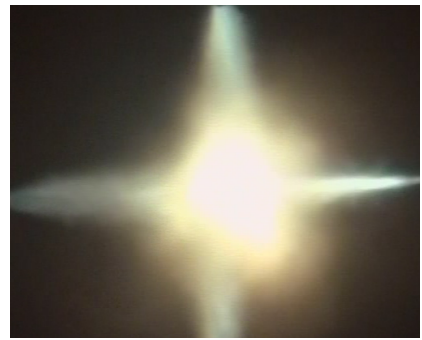}
\caption{(Color online) The typical four-burner impinging flame
image} \label{Fig4}
\end{figure}

A typical flame image is illustrated in Fig.~\ref{Fig4}. The image
analysis included the following steps. First, the video signal was
digitized into 8-bit (256 gray values for each pixel)
two-dimensional digital color images at a rate of 24 frames per
second. Second, for each experimental condition, 50 images were
analyzed, each of them was converted into grey images with the level
ranging from 0 (black) to 255 (white). Third, in order to minimize
the possible generated effects of the gasifier wall background, we
cropped the grey images to remove most of the gasifier wall
background but preserved the whole flame resulting in 200-by-200
images. Four, multifractal analysis was then performed on the 50
resultant images.

\section{Results and discussion}
\label{s1:results}

\subsection{Annealed multifractal DFA of images}

\begin{figure}[htp]
\centering
\includegraphics[width=6.5cm]{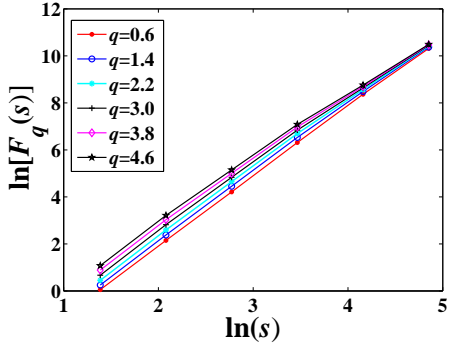}
\caption{The annealed average of detrended fluctuations $F_q(s)$ as
a function of the scale $s$ for six different values of $q$}
\label{Fig5}
\end{figure}

For each experiment, we performed annealed multifractal DFA on 50
arbitrarily chosen flame images. Fig.~\ref{Fig5} shows the
dependence of the annealed average of the detrended fluctuations
$\ln\left[F_q(s)\right]$ with respect to the scale $s$ for six
different orders $q$. The nice linearity of the lines indicates
power-law scaling between $F_q(s)$ and $s$. The scaling range spans
about 2.15 orders of magnitude. According to
\citet{Malcai-Lidar-Biham-Avnir-1997-PRE} and
\citet{Avnir-Biham-Lidar-Malcai-1998-Science}, the scaling range
width for experimental fractality is about 0.5 to 2.0 orders of
magnitude. It means that the power-law scaling observed in our
experiments are quite sound.

\begin{figure}[htp]
\centering
\includegraphics[width=6.5cm]{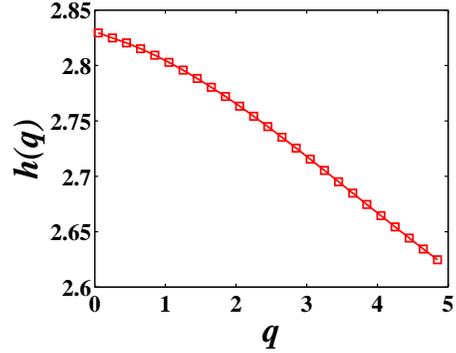}
\caption{Dependence of $h(q)$ with respect to $q$} \label{Fig6}
\end{figure}

According to Eq.~(\ref{Eq:eq7}), the slopes of the straight lines
obtained by least squares regression of $\ln[F_q(s)]$ against
$\ln(s)$ in Fig.~\ref{Fig5} give the estimates of the scaling
exponent $h(q)$. In Fig.~\ref{Fig6} is illustrated $h(q)$ as a
function of $q$ for $0 < q \leqslant 5$. We observe that $h(q)$ is a
nonlinear function of the order $q$, which is the hallmark of
multifractality in the flame images. We note that the estimate of
$F_q(s)$ becomes statistically less significant for larger $q$ due
to the finite size of the flame images. It is also worth stressing
that, no evidence of power-law behaviour is observed for $q\leq0$.
In other words, the scale invariance is destroyed for negative $q$.
This phenomenon is not unusual in the multifractal analysis of
experimental results or natural measures. Examples include the
growth probabilities at perimeter sites of diffusion-limited
aggregations (DLA) \citep{Lee-Stanley-1988-PRL}, the spatial
distribution of the secondary electrons on solid surfaces and in the
bulk \citep{Li-Ding-Wu-1995-PRB,Li-Ding-Wu-1996-PRB}, the growth
probability of a solid-on-solid model \citep{Wang-Wang-Wu-1995-SSC},
the TEM images of four-layered GeAl film after laser irradiation
\citep{Sanchez-Serna-Catalina-Afonso-1992-PRB}, to list a few.

\begin{figure}[htp]
\begin{center}
\includegraphics[width=6.5cm]{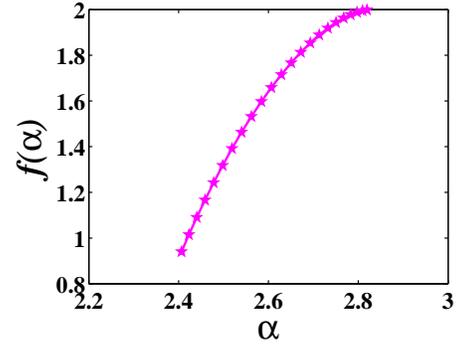}
\caption{Multifractal spectrum for the flame images} \label{Fig7}
\end{center}
\end{figure}

The values of $\alpha(q)$ and $f(\alpha)$ can be computed
numerically based on the Legendre transform
(\ref{Eq:eq9},\ref{Eq:eq10}). The resulting multifractal curve
$f(\alpha)$ is plotted in Fig.~\ref{Fig7} with respect to $\alpha$.
Two fundamental quantities $\alpha_{\min}$ and $\alpha_{\max}$ are
determined, which characterize respectively the minimal and maximal
singular sites of the turbulent flames. Moreover, the width of
singularity strength $\Delta\alpha=\alpha_{\max}-\alpha_{\min}$ can
be used as a measure of heterogeneity of the singular flames.
Similarly, $f(\alpha_{\min})$, $f(\alpha_{\max})$, and
$\Delta{f}=f(\alpha_{\max})-f(\alpha_{\min})$ can also be used to
quantitatively characterize the multifractal nature of the flames.

\subsection{Relationship between multifractal parameters and experimental conditions}

In our experiments, we have investigated 95 conditions by varying
the burner exit velocities of diesel and oxygen,
$M({\rm{m}}/{\rm{s}})$ and $V ({\rm{m}}^3/{\rm{s}})$. Under each
experimental condition, 50 flame images have been used to calculate
the ensemble multifractal parameters $\alpha_{\min}$,
$\alpha_{\max}$, $\Delta\alpha$, $f(\alpha_{\min})$,
$f(\alpha_{\max})$, and $\Delta{f}$. Here we investigate the
possible dependence of these multifractal parameters with respect to
experimental parameters $M$ and $V$ to identify characteristic
multifractal parameters. Specifically, we adopt linear models as
follows
\begin{equation}\label{Eq:Models}
    {\rm{MF}} = \beta_0 + \beta_1 M + \beta_2 V~,
\end{equation}
where ${\rm{MF}}$ stands for the six individual multifractal
parameters. We investigate three different types of models by posing
$\beta_1=0$ or $\beta_2=0$ or freeing $\beta_1$ and $\beta_2$. This
gives 18 models in total. For each model, we use the F-test to check
if the model is statistically significant. If the model is
significant, we employ further the t-test to see whether the model
coefficients are significantly different from zero or not.

As a first step, we consider the six monovariate models with
$\beta_1=0$. The F-tests show that the four models for
$\Delta\alpha$, $f(\alpha_{\min})$, $f(\alpha_{\max})$ and
$\Delta{f}$ are not significant, where all the $p$-values are
greater than 13\%. On the other hand, the $p$-values for
$\alpha_{\min}$ and $\alpha_{\max}$ are both 0.1\%. This implies
that both the multifractal parameters $\alpha_{\min}$ and
$\alpha_{\max}$ are linearly dependent on the burner exit velocity
$V$ of oxygen. In addition, according to the t-tests, we find that
the $p$-values of $\beta_0$ and $\beta_2$ in these two models are
not greater than 0.1\%. In other words, the coefficients $\beta_0$
and $\beta_2$ are significantly different from zero.

Then, we investigate the six monovariate models with $\beta_2=0$.
The F-tests show that the five models for $\alpha_{\min}$,
$\Delta\alpha$, $f(\alpha_{\min})$, $f(\alpha_{\max})$ and
$\Delta{f}$ are not significant, where all the $p$-values are
greater than 29\%, while the $p$-values for $\alpha_{\max}$ is
4.5\%. This means that the multifractal parameter $\alpha_{\max}$ is
linearly dependent on the burner exit velocity $M$ of diesel,
significant at the 5\% level. In addition, according to the t-test,
we find that the $p$-values of $\beta_0$ and $\beta_1$ in this model
are not greater than 5\%. In other words, the coefficients $\beta_0$
and $\beta_1$ are significantly different from zero.

Now we turn to study the six bivariate models (\ref{Eq:Models}) with
different dependent variables ${\rm{MF}}$. The resluts are listed in
Table \ref{TB:Model}. According to the F-tests, the four models for
$\Delta\alpha$, $f(\alpha_{\min})$, $f(\alpha_{\max})$ and
$\Delta{f}$ are not significant whose $p$-values are greater than
14\%. For $\alpha_{\min}$ and $\alpha_{\max}$, the $p$-values
obtained from the F-tests are less than 1\%. Therefore, both the
multifractal parameters $\alpha_{\min}$ and $\alpha_{\max}$ are
linearly dependent on the velocities of oxygen and diesel. However,
the t-tests shows that the coefficients $\beta_1$ for both models
are not significantly different from zero. Speaking differently, the
multifractal parameters $\alpha_{\min}$ and $\alpha_{\max}$ depend
strong upon the burner exit velocity of oxygen and weakly on the
velocity of diesel. Furthermore, the minimal and maximal singularity
strengthes $\alpha_{\min}$ and $\alpha_{\max}$ are characteristic
multifractal parameters that can be used to quantify the
multifractal nature of the turbulence flames.

\begin{table}[htbp]
\centering
    \caption{Identification of characteristic multifractal parameters using linear model (\ref{Eq:Models})}
    \medskip
    \label{TB:Model}
    \begin{tabular}{cccccccccccccccccc}
    \hline\hline
    & \multicolumn{3}{@{\extracolsep\fill}c}{Coefficients} &
    & \multicolumn{3}{@{\extracolsep\fill}c}{t-test} &
    & \multicolumn{2}{@{\extracolsep\fill}c}{F-test}  \\
    \cline{2-4} \cline{6-8} \cline{10-11}
        ${\rm{MF}}$ & $\beta_0$ & $\beta_1$ & $\beta_2$ & %
        & $p_0$ & $p_1$ & $p_2$ && $F$ & $p$ \\\hline %
   $\alpha_{\min}$ & 2.4734 & 0.4538 & -0.0027 && 0.000 & 0.589 & 0.001 & & 5.576 & 0.005 \\%
   $\alpha_{\max}$ & 3.1728 & -0.6503 & -0.0013 && 0.000 & 0.117 & 0.002 & & 7.477 & 0.001 \\%
   $f(\alpha_{\min})$ & 0.3067 & 2.1950 & -0.0011 && 0.649 & 0.260 & 0.555 & & 0.729 & 0.485 \\%
   $f(\alpha_{\max})$ & 1.9911 & -0.0576 & 0.0002 && 0.000 & 0.610 & 0.118 & & 1.279 & 0.283 \\%
   $\Delta\alpha$ & 0.6992 & -1.1035 & 0.0014 && 0.023 & 0.208 & 0.094 & & 1.942 & 0.149 \\%
   $\Delta f$ & 1.6845 & -2.2529 & 0.0013 && 0.013 & 0.243 & 0.491 & & 0.822 & 0.443 \\%
   \hline\hline
   \end{tabular}
\end{table}

\subsection{Discussion}

We have confirmed that the impinging flame images exhibit
multifractal properties. A natural question arises asking which
processes lead to such multifractality. Several examples taken from
other fields in literature provide clues for addressing this
question. \citet{Godano-Alonzo-Vilardo-1997-PAG} argued that the
multifractal nature of the temporal clustering of earthquakes can be
interpreted in terms of diffusive processes of stress in the Earth's
crust. In order to explain the multifractal properties in rainfall
data, \citet{Olsson-Niemczynowicz-1996-JH} made an assumption that a
large-scale flux is successively broken into smaller and smaller
ones in cascades, each receiving an amount of the total flux
specified by a multiplicative process. \citet{Lee-2002-WASP} and
\citet{Lee-Ho-Yu-Wang-Hsiao-2003-Environmetrics} also employed a
random multiplicative process to understand the multifractal
characteristics in air pollutant concentration time series. In the
present case, we submit that a possible explanation for the physical
origin of multifractality in flame images is the atomization and
impinging processes.

Due to the complexity of the atomization process, it is difficult to
clearly describe the mechanism and also impossible to combine all
the influencing factors into one model, such as the equipment
dimensions, the size and geometry of burner, the physical properties
of the dispersed phase and the continuous phase, and the operating
mode. Indeed, \citet{Zhou-Zhao-Wu-Yu-2000-CEJ} proposed a stochastic
multiplicative cascade model for drop breakdown in atomization
processes, which works well in the prediction of drop size
distribution \citep{Liu-Gong-Li-Wang-Yu-2006-CES}. Moreover,
\citet{Zhou-Yu-2001-PRE} studied the multifractal nature of drop
breakup in the air-blast burner atomization process. They applied
the multiplier method to extract the negative and the positive parts
of the $f(\alpha)$ curve with the data of drop-size distribution
measured using dual particle dynamic analyzer. They proposed a
random multifractal model with the multiplier triangularly
distributed to characterize the breakup of drops, the agreement of
the left part ($q>0$) of the multifractal spectrum between the
experimental result and the model is remarkable. Hence, the
breakdown of diesel drops in the gasifier follows a multifractal
process.

Four equal suspension streams flow against one another at high
velocity ($>35~$m/s) and impinge at the center of the gasifier,
resulting in a highly turbulent zone. The gas flows decrease their
axial velocity down to zero at the impingement plane and then
disperse radially, while particles penetrate back and forth between
the opposed streams driven by inertia and friction forces. This
impinging process leads to locally singular distributions of the raw
materials.

\section{Conclusion}
\label{s1:Concl}

The impinging process has proved to be one of the most effective
methods enhancing heat and mass transfer in multiphase environment.
On a laboratory-scale testing platform of impinging entrained-flow
gasifier with four opposed burners, the flame images for diesel
combustion and gasification process were measured with a single
charge coupled device (CCD) camera. Ninety-five experimental
conditions were investigated.

The multifractal properties of the turbulent flames have been
investigated using the two-dimensional multifractal detrended
fluctuation analysis, which is accurate and easy to implement for
image analysis. Nice power-law scaling is unveiled in the annealed
average of detrended fluctuations when the order $q>0$. The scaling
exponent $h(q)$ is a nonlinear function of $q$, which confirms the
multifractal nature of the flames under investigation. We argue that
the multifractality in gasification flames stems from the
multiplicative process of atomization and the impinging process of
raw materials.

We analyzed the relationship between six multifractal parameters
($\alpha_{\min}$, $\alpha_{\max}$, $\Delta\alpha$,
$f(\alpha_{\min})$, $f(\alpha_{\max})$ and $\Delta{f}$) and the
velocities of oxygen and diesel ($V$ and $M$). Two multifractal
parameters $\alpha_{\min}$ and $\alpha_{\max}$ have been identified
by extensive F-tests as characteristics of the observed multifractal
nature, which are dependent on the velocities linearly. The t-tests
show that the velocity of oxygen has greater impact on the
multifractality of flames. These analyses enable us to gain a better
understanding of the complexity of the combustion dynamics of
four-burner impinging entrained-flow gasification.

\bigskip{\textbf{Acknowledgments:}}

We are grateful to Gao-Feng Gu for useful discussions. This work was
partially supported by the National Basic Research Program of China
(No. 2004CB217703), the PCSIRT (IRT0620), the Program for New
Century Excellent Talents in University (NCET-05-0413), and the
Project Sponsored by the Scientific Research Foundation for the
Returned Overseas Chinese Scholars, State Education Ministry.

\bigskip

\begin{table}[h]
  \flushleft
  \begin{tabular}{lll}
    \noindent{\large\textbf{Notations}}&\\
    $s$                                  & scale of boxes \\
    $T=50$                               & number of images\\
    $X(i,j)$                             & two-dimensional array \\
    $u_{\rm v,w}(t;i,j)$                 & cumulative sum for image $t$\\
    $\widetilde{u}_{v,w}(i,j)$           & fitting bivariate polynomial \\
    $\epsilon_{v,w}(t;i,j)$              & residual matrix \\
    $q$                                  & order of detrended fluctuation function\\
    $F_{q}(s)$                           & detrended fluctuation function \\
    $D_{f}$                              & fractal dimension \\
    $h(q)$                               & scaling exponent function \\
    $\tau(q)$                            & mass exponent function \\
    $\alpha(q)$                          & singularity strength function \\
    $\alpha_{\rm max}$                   & maximum singularity \\
    $\alpha_{\rm min}$                   & minimum singularity \\
    $\Delta \alpha$                      & width of multifractal spectrum\\
    $f(\alpha)$                          & multifractal singularity spectrum    \\
    $\Delta f$                           & difference, $\Delta f=f(\alpha_{\max})-f(\alpha_{\min})$ \\
    $M$                                  & burner exit velocity of diesel (m/s)\\
    $V$                                  & burner exit velocity of oxygen (m/s) \\
    $\beta_0, \beta_1, \beta_2$          & model coefficients \\
    $p_0, p_1, p_2$                      & $p$-values of model coefficients \\
  \end{tabular}
\end{table}

\bibliography{E:/Papers/Auxiliary/Bibliography_FullJournal} 

\begin{thebibliography}{42}
\expandafter\ifx\csname natexlab\endcsname\relax\def\natexlab#1{#1}\fi
\expandafter\ifx\csname url\endcsname\relax
  \def\url#1{\texttt{#1}}\fi
\expandafter\ifx\csname urlprefix\endcsname\relax\def\urlprefix{URL }\fi

\bibitem[{Avnir et~al.(1998)Avnir, Biham, Lidar, and
  Malcai}]{Avnir-Biham-Lidar-Malcai-1998-Science}
Avnir, D., Biham, O., Lidar, D., Malcai, O., 1998. {Is the geometry of nature
  fractal?} Science 279, 39--40.

\bibitem[{Berman and Tamir(1996)}]{Berman-Tamir-1996-CJCE}
Berman, Y., Tamir, A., 1996. {Experimental investigation of phosphate dust
  collection in impinging streams (IS)}. Canadian Journal of Chemical
  Engineering 74, 817--821.

\bibitem[{Berman et~al.(2000{\natexlab{a}})Berman, Tanklevsky, Oren, and
  Tamir}]{Berman-Tanklevsky-Oren-Tamir-2000a-CES}
Berman, Y., Tanklevsky, A., Oren, Y., Tamir, A., 2000{\natexlab{a}}. {Modeling
  and experimental studies of $\rm SO_2$ absorption in coaxial cylinders with
  impinging streams: Part I}. Chemical Engineering Science 55, 1009--1021.

\bibitem[{Berman et~al.(2000{\natexlab{b}})Berman, Tanklevsky, Oren, and
  Tamir}]{Berman-Tanklevsky-Oren-Tamir-2000b-CES}
Berman, Y., Tanklevsky, A., Oren, Y., Tamir, A., 2000{\natexlab{b}}. {Modeling
  and experimental studies of $\rm SO_2$ absorption in coaxial cylinders with
  impinging streams: part II}. Chemical Engineering Science 55, 1023--1028.

\bibitem[{Champion and Libby(1993)}]{Champion-Libby-1993-PFA}
Champion, M., Libby, P.~A., 1993. {Reynolds stress description of opposed and
  impinging turbulent jets. Part I: Closely spaced opposed jets}. Physics of
  Fluids A 5, 203--216.

\bibitem[{Das and Evans(1997)}]{Das-Evans-1997-EF}
Das, A.~K., Evans, R.~L., 1997. {An experimental study to determine fractal
  parameters for lean premixed flames}. Experiments in Fluids 22, 312--320.

\bibitem[{Dehkordi(2002)}]{Dehkordi-2002-CEP}
Dehkordi, A.~M., 2002. {Application of a novel-opposed-jets contacting device
  in liquid-liquid extraction}. Chemical Engineering \& Processing 41,
  251--258.

\bibitem[{Erard et~al.(1996)Erard, Boukhalfa, Puechberty, and
  Trinit{\'e}}]{Erard-Boukhalfa-Puechberty-Trinite-1996-CST}
Erard, V., Boukhalfa, A., Puechberty, D., Trinit{\'e}, M., 1996. {A statistical
  study on surface properties of freely-propagating premixed turbulent flames}.
  Combustion Science and Technology 113-114, 313--327.

\bibitem[{Godano et~al.(1997)Godano, Alonzo, and
  Vilardo}]{Godano-Alonzo-Vilardo-1997-PAG}
Godano, C., Alonzo, M.~L., Vilardo, G., 1997. {Multifractal approach to time
  clustering of earthquakes application to Mt. Vesuvio seismicity}. Pure and
  Applied Geophysics 149, 375--390.

\bibitem[{Goix and Shepherd(1993)}]{Goix-Shepherd-1993-CST}
Goix, P., Shepherd, I.~G., 1993. {Lewis number effects on turbulent premixed
  flame structure}. Combustion Science and Technology 91, 191--206.

\bibitem[{Goix et~al.(1989)Goix, Shepherd, and
  Trinit{\'e}}]{Goix-Shepherd-Trinite-1989-CST}
Goix, P.~J., Shepherd, I.~G., Trinit{\'e}, M., 1989. {A fractal study of a
  premixed V-shaped $\rm H_2$/air flame}. Combustion Science and Technology 63,
  275--286.

\bibitem[{Gouldin(1987)}]{Gouldin-1987-CF}
Gouldin, F.~C., 1987. {An application of fractals to modeling premixed
  turbulent flames}. Combustion and Flame 68, 249--266.

\bibitem[{Gouldin et~al.(1989)Gouldin, Hilton, and
  Lamb}]{Gouldin-Hilton-Lamb-1989-PCI}
Gouldin, F.~C., Hilton, S.~M., Lamb, T., 1989. {Experimental evaluation of the
  fractal geometry of flamelets}. Symposium (International) on Combustion 22,
  541--550.

\bibitem[{Gu and Zhou(2006)}]{Gu-Zhou-2006-PRE}
Gu, G.-F., Zhou, W.-X., 2006. {Detrended fluctuation analysis for fractals and
  multifractals in higher dimensions}. Physical Review E 74, 061104.

\bibitem[{Halsey et~al.(1986)Halsey, Jensen, Kadanoff, Procaccia, and
  Shraiman}]{Halsey-Jensen-Kadanoff-Procaccia-Shraiman-1986-PRA}
Halsey, T.~C., Jensen, M.~H., Kadanoff, L.~P., Procaccia, I., Shraiman, B.~I.,
  1986. {Fractal measures and their singularities: The characterization of
  strange sets}. Physical Review A 33, 1141--1151.

\bibitem[{Kantelhardt et~al.(2002)Kantelhardt, Zschiegner, Koscielny-Bunde,
  Havlin, Bunde, and
  Stanley}]{Kantelhardt-Zschiegner-Bunde-Havlin-Bunde-Stanley-2002-PA}
Kantelhardt, J.~W., Zschiegner, S.~A., Koscielny-Bunde, E., Havlin, S., Bunde,
  A., Stanley, H.~E., 2002. {Multifractal detrended fluctuation analysis of
  nonstationary time series}. Physica A 316, 87--114.

\bibitem[{Kostiuk and Libby(1993)}]{Kostiuk-Libby-1993-PFA}
Kostiuk, L.~W., Libby, P.~A., 1993. {Comparison between theory and experiment
  for turbulence in opposed streams}. Physics of Fluids A 5, 2301--2303.

\bibitem[{Lee(2002)}]{Lee-2002-WASP}
Lee, C.~K., 2002. {Multifractal characteristics in air pollutant concentration
  time series.} Water, Air, and Soil Pollution 135, 389--409.

\bibitem[{Lee et~al.(2003)Lee, Ho, Yu, Wang, and
  Hsiao}]{Lee-Ho-Yu-Wang-Hsiao-2003-Environmetrics}
Lee, C.~K., Ho, D.~S., Yu, C.~C., Wang, C.~C., Hsiao, Y.~H., 2003. {Simple
  multifractal cascade model for the air pollutant concentration time series.}
  Environmetrics 14, 255--269.

\bibitem[{Lee and Stanley(1988)}]{Lee-Stanley-1988-PRL}
Lee, J., Stanley, H.~E., 1988. {Phase transition in the multifractal spectrum
  of diffusion-limited aggregation}. Physical Review Letters 61, 2945--2948.

\bibitem[{Li et~al.(1995)Li, Ding, and Wu}]{Li-Ding-Wu-1995-PRB}
Li, H., Ding, Z.-J., Wu, Z.-Q., 1995. {Multifractal behavior of the
  distribution of secondary emission sites on solid surfaces}. Physical Review
  B 51, 13554--13559.

\bibitem[{Li et~al.(1996)Li, Ding, and Wu}]{Li-Ding-Wu-1996-PRB}
Li, H., Ding, Z.-J., Wu, Z.-Q., 1996. {Multifractal analysis of the spatial
  distribution of secondary-electron emission sites}. Physical Review B 53,
  16631--16636.

\bibitem[{Liu et~al.(2006)Liu, Gong, Li, Wang, and
  Yu}]{Liu-Gong-Li-Wang-Yu-2006-CES}
Liu, H.-F., Gong, X., Li, W.-F., Wang, F.-C., Yu, Z.-H., 2006. {Prediction of
  droplet size distribution in sprays of prefilming air-blast atomizers}.
  Chemical Engineering Science 61, 1741--1747.

\bibitem[{Malcai et~al.(1997)Malcai, Lidar, Biham, and
  Avnir}]{Malcai-Lidar-Biham-Avnir-1997-PRE}
Malcai, O., Lidar, D.~A., Biham, O., Avnir, D., 1997. {Scaling range and
  cutoffs in empirical fractals}. Physical Review E 56, 2817--2828.

\bibitem[{Mandelbrot(1983)}]{Mandelbrot-1983}
Mandelbrot, B.~B., 1983. {The Fractal Geometry of Nature}. W. H. Freeman, New
  York.

\bibitem[{Mantzaras et~al.(1989)Mantzaras, Felton, and
  Bracco}]{Mantzaras-Felton-Bracco-1989-CF}
Mantzaras, J., Felton, P.~G., Bracco, F.~V., 1989. Fractals and turbulent
  premixed engine flames. Combustion and Flame 77, 295--310.

\bibitem[{Murayama and Takeno(1988)}]{Murayama-Takeno-1989-PCI}
Murayama, M., Takeno, T., 1988. {Fractal-like character of flamelets in
  turbulent premixed combustion}. Symposium (International) on Combustion 22,
  551--559.

\bibitem[{North and Santavicca(1990)}]{North-Santavicca-1990-CST}
North, G.~L., Santavicca, D.~A., 1990. {The fractal nature of premixed
  turbulent flames}. Combustion Science and Technology 72, 215--232.

\bibitem[{Nosseir and Behart(1986)}]{Nosseir-Behart-1986-AIAAJ}
Nosseir, N.~S., Behart, S., 1986. {Characteristics of jet impingement in a
  side-dump combustor}. AIAA Journal 24, 1752--1757.

\bibitem[{Olsson and Niemczynowicz(1996)}]{Olsson-Niemczynowicz-1996-JH}
Olsson, J., Niemczynowicz, J., 1996. {Multifractal analysis of daily spatial
  rainfall distributions}. Journal of Hydrology 187, 29--43.

\bibitem[{Peng et~al.(1994)Peng, Buldyrev, Havlin, Simons, Stanley, and
  Goldberger}]{Peng-Buldyrev-Havlin-Simons-Stanley-Goldberger-1994-PRE}
Peng, C.-K., Buldyrev, S.~V., Havlin, S., Simons, M., Stanley, H.~E.,
  Goldberger, A.~L., 1994. {Mosaic organization of DNA nucleotides}. Physical
  Review E 49, 1685--1689.

\bibitem[{Sanchez et~al.(1992)Sanchez, Serna, Catalina, and
  Afonso}]{Sanchez-Serna-Catalina-Afonso-1992-PRB}
Sanchez, A., Serna, R., Catalina, F., Afonso, C., 1992. {Multifractal patterns
  formed by laser irradiation in GeAl thin multilayer films}. Physical Review B
  46, 487--490.

\bibitem[{Shepherd et~al.(1991)Shepherd, Cheng, and
  Goix}]{Shepherd-Cheng-Goix-1991-PCI}
Shepherd, I.~G., Cheng, R.~K., Goix, P.~J., 1991. {The spatial scalar structure
  of premixed turbulent stagnation point flames}. Symposium (International) on
  Combustion 23, 781--787.

\bibitem[{Smallwood et~al.(1995)Smallwood, G{\"u}lder, Snelling, Deschamps, and
  G{\"o}kalp}]{Smallwood-Gulder-Snelling-Deschamps-Gokalp-1995-CF}
Smallwood, G.~J., G{\"u}lder, {\"O}.~L., Snelling, D.~R., Deschamps, B.~M.,
  G{\"o}kalp, I., 1995. {Characterization of flame front surfaces in turbulent
  premixed methane/air combustion}. Combustion and Flame 101, 461--470.

\bibitem[{Tamir(1994)}]{Tamir-1994}
Tamir, A., 1994. {Impinging-Stream Reactors: Fundamentals and Applications}.
  Elsevier, Amsterdam.

\bibitem[{Tamir et~al.(1984)Tamir, Elperin, and
  Luzzatto}]{Tamir-Elperin-Luzzatto-1984-CES}
Tamir, A., Elperin, I., Luzzatto, K., 1984. {Drying in a new two impinging
  streams reactor}. Chemical Engineering Science 39, 139--146.

\bibitem[{Wang et~al.(1995)Wang, Wang, and Wu}]{Wang-Wang-Wu-1995-SSC}
Wang, B., Wang, Y., Wu, Z.-Q., 1995. {Multifractal behavior of solid-on-solid
  growth}. Solid State Communications 96, 69--72.

\bibitem[{Wu et~al.(1991)Wu, Kwon, Driscoll, and
  Faeth}]{Wu-Kwon-Driscoll-Faeth-1991-CST}
Wu, M.~S., Kwon, S., Driscoll, J.~F., Faeth, G.~M., 1991. {Preferential
  diffusion effects on the surface stucture of turbulent premixed hydrogen/air
  flames}. Combustion Science and Technology 78, 69--96.

\bibitem[{Yoshida et~al.(1994{\natexlab{a}})Yoshida, Ando, Yanagisawa, and
  Tsuji}]{Yoshida-Ando-Yanagisawa-Tsuji-1994-CST}
Yoshida, A., Ando, Y., Yanagisawa, T., Tsuji, H., 1994{\natexlab{a}}. {Fractal
  behavior of wrinkled laminar flame}. Combustion Science and Technology 96,
  121--134.

\bibitem[{Yoshida et~al.(1994{\natexlab{b}})Yoshida, Kasahara, and
  Yanagisawa}]{Yoshida-Kasahara-Tsuji-Yanagisawa-1994-CST}
Yoshida, A., Kasahara, M.~Tsuji, H., Yanagisawa, T., 1994{\natexlab{b}}.
  {Fractal geometry application in estimation of turbulent burning velocity of
  wrinkled laminar flame}. Combustion Science and Technology 103, 207--218.

\bibitem[{Zhou and Yu(2001)}]{Zhou-Yu-2001-PRE}
Zhou, W.-X., Yu, Z.-H., 2001. {Multifractality of drop breakup in the air-blast
  nozzle atomization process}. Physical Review E 63, 016302.

\bibitem[{Zhou et~al.(2000)Zhou, Zhao, Wu, and Yu}]{Zhou-Zhao-Wu-Yu-2000-CEJ}
Zhou, W.-X., Zhao, T.-J., Wu, T., Yu, Z.-H., 2000. {Application of fractal
  geometry to atomization process}. Chemical Engineering Journal 78, 193--197.

\end{thebibliography}


\end{document}